# An inorganic ABX$_3$ perovskite materials dataset for target property prediction and classification using machine learning

Ericsson Tetteh Chenebuah [†, *] and David Tetteh Chenebuah [‡]

[†] Department of Mechanical Engineering, University of Ottawa

[‡] Department of Metallurgical and Materials Engineering, University of Nigeria

*echen013@uottawa.ca

**Abstract**

The reliability with Machine Learning (ML) techniques in novel materials discovery often depend on the quality of the dataset, in addition to the relevant features used in describing the material. In this regard, the current study presents and validates a newly processed materials dataset that can be utilized for benchmark ML analysis, as it relates to the prediction and classification of deterministic target properties. Originally, the dataset was extracted from the Open Quantum Materials Database (OQMD) and contains a robust 16,323 samples of ABX$_3$ inorganic perovskite structures. The dataset is tabular in form and is preprocessed to include sixty-one generalized input features that broadly describes the physicochemical, stability/geometrical, and Density Functional Theory (DFT) target properties associated with the elemental ionic sites in a three-dimensional ABX$_3$ polyhedral. For validation, four different ML models are employed to predict three distinctive target properties, namely: formation energy, energy band gap, and crystal system. On experimentation, the best accuracy measurements are reported at 0.013 eV/atom MAE, 0.216 eV MAE, and 85% F1, corresponding to the formation energy prediction, band gap prediction and crystal system multi-classification, respectively. Moreover, the realized results are compared with previous literature and as such, affirms the resourcefulness of the current dataset for future benchmark materials analysis via ML techniques. The preprocessed dataset and source codes are openly available to download from github.com/chenebuah/ML_abx3_dataset.

1. **Introduction**

It is unsurprising that inorganic perovskite structures are at the forefront of novel energy materials discovery thanks to their vast compositional and configurational setup, which makes them attractive in diverse engineering applications. Among the several types of perovskite structures is the $ABX_3$ stoichiometry consisting predominantly of three chemical elements that span across the entire periodic table. Ideally, the B-site metallic cation occupies the center of a corner sharing octahedron and is coordinated by six anionic X-site elements, whereas the A-site metallic cation is situated within a twelve-fold cavity of a three-dimensional polyhedral geometry [1]. Owing to its chemical ubiquitousness, the $ABX_3$ perovskite structure is known to exhibit certain multifunctionality extending to superconductivity, piezoelectricity, ferroelectricity, optoelectricity, catalysis, and so on. The multifunctionality afforded with perovskites is often tied to the desirable properties inherent to their crystal structure (i.e. structure → property relationship). For instance, perovskite solar conversion cells are widely harnessed in the photovoltaic industry, which is because of their tunable band gaps for optimally matching specific wavelengths emitted from solar radiation [2]. Hence, accurately estimating perovskite functional/target properties is of great significance to the materials science community due to the role that property may play in defining the specific application. Inasmuch as there could be several other mechanical and electronic properties of compelling interest to researchers, the formation energy, band gap, and crystal system are highlighted as critical in characterizing the usability of general crystalline energy materials. By definition, the band gap is the energy difference between the lowest unoccupied conduction band and highest occupied valence band at the fermi level, and assists in classifying the electronic state of the material as either conductive, semi-conductive, or insulative [3]. The formation energy defines the potential formability of a material with respect to disintegration of constitutive atoms that form the bulk crystal material, and is required to generate phase diagrams [4]. The crystal system on the other hand plays a unique role in characterizing the geometrical (stable) and physical morphology of the material [5]. Conventionally, the aforementioned properties are proven to be determined via first-principles deterministic methods and/or experimental synthesization. Despite the immense contribution of conventional approaches in the past and present, they are known to be unpractical for extremely wide materials design search space. For instance, first-principles techniques such as Density Functional Theory (DFT) are computationally expensive due to their derived solutions from solving an approximation of the Schrödinger equation whereas synthesization are suggested to Edisonian-based

(i.e. trial-and-error) and resource demanding. On this frontier, Machine Learning (ML) techniques are widely revered to be quicker, efficient and inexpensive unlike traditional DFT and Edisonian experiments. However, the reliability with ML critically hinges on the validity, clarity and acceptability of data as such qualities are beneficial for accurate model training. This, therefore, explains why several well-established databases in solid-state materials are devoted towards the continuous improvement of their data platforms in integrating new material entries with computed target properties that are of close approximation to actual experiments. Some proven materials data platforms include the Materials Project (MP as driven by *pymatgen*) [6] and Open Quantum Materials Database (OQMD) [7]. Both MP and OQMD accommodate bulk experimental and DFT stored data, and are open-sourced, thereby enabling unrestricted access for material data informatics via ML. Prior to imminent ML investigation, a user is required to preprocess the data afforded by these platforms and further feature engineer them using specially customized materials descriptors. As an example, Table 1 outlines some benchmark ML investigations that uses preprocessed training data for target property prediction and classification. The table also reveals their reported ML evaluation scores, as comprehensively investigated on formation energy prediction, band gap prediction, and crystal system multi-classification. For example, a unique (i.e. non-polymorphic) perovskite tabular-dataset consisting of about 1,308 training samples was harnessed from the Materials Project (MP) in a previous study, and was used to train 12 ML models. From the investigation, the best Mean Absolute Error (MAE) predictive scores were obtained at 0.055 eV/atom and 0.462 eV for formation energy and band gap, respectively [8]. Moreover, in a different study on crystal system multi-classification, 675 preprocessed oxide-perovskite compounds were ML classified into their respective crystal systems with performance evaluation reported at 80.3% in accuracy using the Light Gradient Boosting algorithm [9]. In another study, 606 newly screened dataset from the materials project and OQMD were conjointly used to predict non-metallic (finite) oxide-perovskite bandgaps with MAE scores realized at 0.384 eV [3]. All aforementioned studies and many more contribute to dataset validation in general, which therefore leads improvement pathways for further benchmark analysis using updated datasets.

Table 1. Examples of recent benchmark evaluation defining the predictive accuracy of formation energy and band gap, and the multi-classification accuracy of crystal systems for general inorganic crystal structures.

| Ref | Prediction/Classification Technique | Evaluation | Train data |
|---|---|---|---|
| **Formation Energy (eV/atom)** | | | |
| [10] | Crystal Graph Convolutional Neural Networks (CGCNN) | 0.039 MAE | 28,046 |
| [8] | Support Vector Regression (SVR) | 0.055 MAE | 1,308 |
| [4] | Generalized Gradient Approximation (GGA-DFT) + U | 0.081 – 0.136 MAE | 1,670 |
| [3] | Bagging | 0.086 MAE | 606 |
| [11] | Decision Forest | 0.088 MAE | 228,676 |
| [12] | Light Gradient Boosting Machine + Efficient Global Optimization (EGO) | 0.160 MAE | 1,250 |
| **Energy Bandgap (eV)** | | | |
| [13] | Random Forest | 0.149 MAE | 432 |
| [14] | Materials Graph Network (MEGNet) | 0.280 MAE | 10,000 |
| [3] | Gradient Boosting Regression (GBR) | 0.384 MAE | 606 |
| [10] | Crystal Graph Convolutional Neural Networks (CGCNN) | 0.388 MAE | 16,458 |
| [15] | eXtreme Gradient Boosting Regression (XGBoost) | 0.319 RMSE | 1,461 |
| [8] | Support Vector Regression (SVR) | 0.462 MAE | 1,308 |
| [16] | Generalized Gradient Approximation (GGA-DFT) + U | 0.6 MAE DFT | 80,000 |
| **Crystal System Classification** | | | |
| [17] | Random Forest | 72.3% - 84.4% | 60,636 |
| [9] | Light Gradient Boosting Machine | 80.30% | 675 |
| [18] | Random Forest | 81.60% | 125,276 |
| [19] | Deep Convoluted Neural Network (DCNN) | 94.90% | 150,000 |

Hence, in the current study, a newly processed perovskite materials tabular-dataset is presented and validated for standardizing benchmark ML analysis, as it relates to target property prediction and classification. The processed dataset is made openly available and is suggested to be resourceful for improving ML predictive capability, as well as fine-tuning models for hyperparameter optimization. The dataset contains a robust 16,323 of $ABX_3$ samples, all originally extracted from the OQMD platform, and captures a diverse mix of experimental and first-principle deterministic possibilities that generally occur with the perovskite crystal structure. For feature engineering, each material in the dataset is described using sixty-one input features, which includes three target properties, namely: formation energy, band gap and crystal system. The dataset is experimented using four highly efficient tabular-dataset ML models, namely: Support Vector Machine (SVM), Random Forest (RFR), eXtreme Gradient Boosting Machine (XGB), and Light Gradient Boosting (LGB) Machine. The modeling performances of all considered model are evaluated and compared with previous literature for benchmarking the reliability of the present dataset.

2. **Dataset generation and preprocessing**

All original data samples utilized in the study were generated and extracted from the Open Quantum Materials Database (OQMD). The OQMD platform is a high-throughput materials database with more than one million density functional theory (DFT) entries in total energy calculations, consisting of both Inorganic Crystal Structure Database (ICSD) experimental structures and hypothetical structures [7]. To accentuate on a modeling process that is material-class specific, the extracted samples considered in the analysis were screened to ensure they satisfy perovskite description. The first screening process was based on eliminating inverse or anti-perovskite inorganic structures [20]. As a result, only common X-anionic elements that are characteristic to the three-dimensional polyhedral setting within a perovskite lattice configuration were selected. Such anionic elements and their common charge states were determined to be $Br^-$, $Cl^-$, $F^-$, $H^-$, $I^-$, $O^{2-}$, $S^{2-}$, $Se^{2-}$, $Te^{2-}$, $P^{3-}$ and $N^{3-}$. Moreover, unlike some ML research that consider only oxide-perovskites in their investigations [3, 9], the presented dataset is expanded to include other anionic forms apart from oxygen, thereby preserving its generalization. In addition to cleaning the samples based on anionic elements, some perovskite structures that were identified as critically unstable were removed. These compounds were labeled as sample outliers that can potentially obscure the ML process. They were characterized by formation energy and/or stability energy with values greater than 5 eV/atom. The data analytical cleaning process accumulated into 16,323 samples consisting of seven crystal systems belonging to the primitive Bravais lattice. Figure 1 describes a frequency-modeled periodic table showing all A- and B- sites chemical cations. As can be observed, the cationic sites consist of a wide mix of possible chemical elements spanning over Groups I–XV of the periodic table, including lanthanides and actinides. Furthermore, Fig. 2(a) visualizes Kernel Density Estimate (KDE) curves for the formation energy and band gap targets, revealing their respective data distributions. Statistically, the newly processed dataset comprise about 79% of compounds that have infinite band gaps (i.e. $Eg = 0$) and 21% that have finite band gaps (i.e. $Eg > 0$). In addition, about 97% of all compounds have negative formation energies (i.e. considered generally to be stable) and 3% have positive formation energies. The value counts of all crystal systems emerging from the screening process are illustrated in Fig. 2(b), showing that cubic, trigonal, orthorhombic and tetragonal crystal systems are well represented in the dataset and constitute over 94% of the overall sample size.

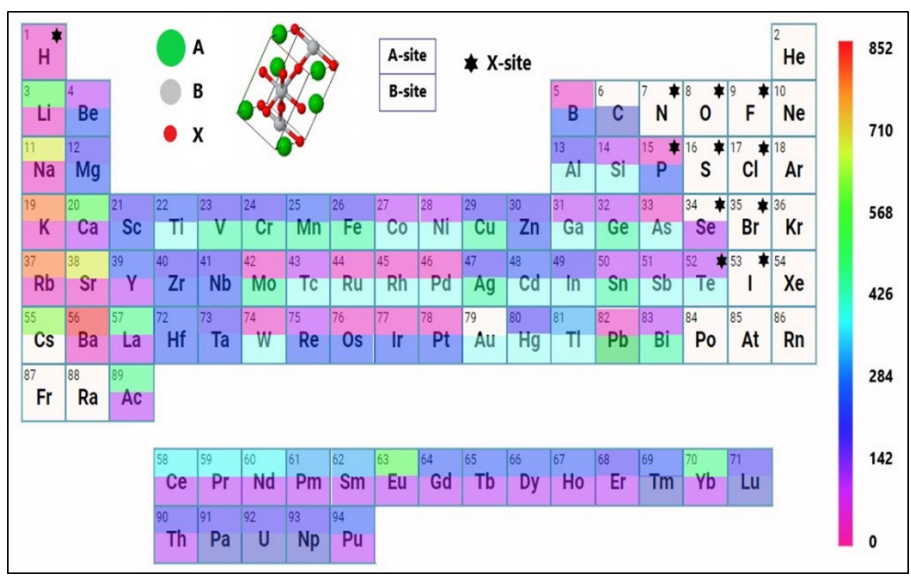

Figure 1. Frequency-modeled periodic table showing A- and B- sites chemical cations, and asterisked X-site anions from the preprocessed OQMD dataset used in the present study.

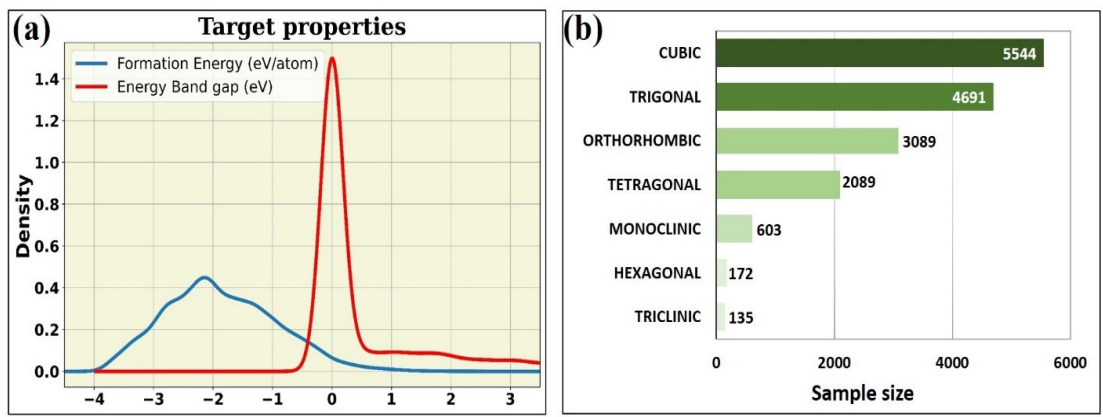

Figure 2. (a) Kernel Density Estimate (KDE) curves revealing frequency distribution of electronic properties in the dataset; (b) Bar chart displaying respective sample sizes for each crystal system.

Table 2. All features considered in study. For each feature, the corresponding abbreviation and unit of measurement are outlined.

| Categories | Features | Abbreviation | Unit | Ref |
|---|---|---|---|---|
| **Physicochemical (mean and standard deviation)** | Atomic Number | $Z$ | - | [6] |
| | Group Number | $grp$ | - | [6] |
| | Row Number | $row$ | - | [6] |
| | Pettifor Mendeleev Number | $pet\_mn$ | - | [23] |
| | Villars Modified Mendeleev Number | $mn$ | - | [24] |
| | IUPAC Ordering Number | $iupac$ | - | [6] |
| | Average Ionic Radius | $av\_ionrad$ | angstrom | [6] |
| | Atomic Radius | $atom\_rad$ | angstrom | [6] |
| | Calculated Atomic Radius | $cal\_atom\_rad$ | angstrom | [6] |
| | Covalent Radius | $cov\_rad$ | angstrom | [25] |
| | Van der Waals Radius | $vdw$ | angstrom | [6] |
| | Pauling Electronegativity | $x$ | pauling | [26] |
| | Electron Affinity | $ea$ | eV | [26] |
| | First Ionization Energy | $ie$ | eV | [26] |
| | Static Average Electric Dipole Polarizability | $p$ | $10^{-24}$ cm$^3$ | [26] |
| | Valence Electrons | $val$ | - | - |
| | Atomic Mass | $atom\_mass$ | g/mol | [6] |
| | Elemental Solid Density | $rho$ | g/cc | [6] |
| | Molar Volume | $mol\_vol$ | cc | [6] |
| | Boiling Point | $bp$ | kelvin | [6] |
| | Melting Point | $mp$ | kelvin | [6] |
| | Thermal Conductivity | $k$ | W/m.k | [6] |
| | Heat of Fusion | $heat\_fus$ | eV | [6] |
| | Heat of Vaporization | $haet\_vap$ | eV | [6] |
| | Specific Heat | $spec\_heat$ | KJ/kg.k | [6] |
| | Sum of sp-Bonding Radius | $rsp$ | atomic unit | [27] |
| | Average of sp-Bonding Radius | $av\_rsp$ | atomic unit | [27] |
| **Stability & Geometrical** | Goldschmidt Tolerance Factor | $gtf$ | - | [21] |
| | Octahedral Factor | $Of$ | - | - |
| **OQMD properties** | Unit Cell Volume | $vol$ | cubic-Angstrom | [7] |
| | Crystal system | $Cs$ | - | [7] |
| | Stability Energy | $Es$ | eV/atom | [7] |
| | Formation Energy | $Ef$ | eV/atom | [7] |
| | Energy Bandgap | $Eg$ | eV | [7] |

3.   **Feature Selection**

The input features utilized in the present study are referred to as generalized compositional features, and are in fact similar to those used in a previous study on target property prediction of $ABX_3$ and $A_2BB'X_6$ perovskite structures [3, 8]. They are broadly divided into three categories that are representative of the functional and stability/geometrical behavior with respect to a lattice-translated perovskite crystal. The first category of features describe the physicochemical properties affiliated to the specific chemical element occupying the distinctive ionic sites within the $ABX_3$ polyhedral, and provide 55 features in total. For each physicochemical property, the ionic feature representations are compressed into their mean and standard deviation. The second category of input features describe the calculated stability criteria based on Goldschmidt tolerance factor and octahedral factor [21]. Finally, the third category is composed of OQMD extracted properties, which includes space group notation, unit cell volume, stability energy, formation energy, and band gap. Because this study deals with the multi-classification of crystal systems, all respective space-group notations are therefore matched to their corresponding crystal structures, appertaining to their notational origin from solid-state symmetrical operations [22]. Moreover, using the unit cell volume (*vol*) as an input feature, additionally serves the purpose of differentiating polymorphic duplicate structures within the dataset, which ensures that each sample is uniquely represented. Overall, 61 features were considered in total, which includes the three core target properties that are subjected to modeling simulation (i.e. *Ef*, *Eg* and *Cs*). Table 2 provides a concise overview of all features considered in study, as described by their aforementioned categories.

4.   **Machine learning Models**

This study applies four ML tabular-dataset models for both regression and classification analysis. They include Support Vector Machine (SVM) [28], Random Forest (RF) [29], eXtreme Gradient Boosting (XGB) [30], and Light Gradient Boosting Machine (LGBM) [31]. These models are appropriately suited for tabular datasets and were chosen based on their previously demonstrated performance for accurately predicting target properties, as related to general inorganic solid-state crystals (Table 1). Depending on the task involved, all model algorithms are conditioned to effect the mode of the predictive experiment (i.e. either Regression or Classification modes), as implemented using Scikit-Learn [32]. As such, the ML models are evaluated using appropriate metrics that specifically relates to the given investigation. For example, considering the regressive analysis of formation energy (*Ef*) and band gap (*Eg*) targets, the metrics used for evaluation are the Mean Absolute Error (MAE), Root Mean Squared Error (RMSE), and coefficient of determination ($R^2$). Similarly, for crystal system multi-classification, average F1-scores (i.e. accuracy), Precision scores, and Recall scores are used to calibrate the modeling performance for each ML model. The tabulated ABX$_3$ feature dataset is therefore formulated into a $M \times N$ matrix, represented as $= [x_{MN}]$, where $x$ is a continuous value/variable that corresponds to the $M^{th}$ perovskite sample of the $N^{th}$ perovskite input feature.

## 5. Results and discussion

In the present study, the results are divided into three subsections. The first subsection presents the formation energy (*Ef*) predictive errors and graphically illustrates the regressive performance fittings for all considered models according to their respective $R^2$ values. Likewise, the second subsection reports the energy band gap (*Eg*) modeling performance and graphs the predictive results for comparative inspection. In the third subsection, the multi-classification modeling experiment of crystal systems (*Cs*) are reported based on downsampling and oversampling (data augmentation) techniques. For each subsection moreover, the dataset is reorganized in terms of the experimented sample size and number of input features used for simulation.

### 5.1. Formation energy (*Ef*) prediction

For predicting the formation energy target, 58 input features were effectively used for training and testing all considered ML models. The features include all physicochemical and stability properties (as outlined using the Table 2), in addition to the unit cell volume (*vol*) and stability energy (*Es*). As a result, the dataset is organized into a $M \times N = 16,323$ (perovskites) $\times$ 58 (features) matrix, and is further split into 70% for training and 30% for testing. As reported using the Table 3, the Support Vector Machine (SVM) is seen to outperform its peers in predicting the formation energy target based on test set evaluation. The accuracy measurements for SVM are reported at 0.013 eV/atom, 0.070 eV/atom, and 99.45% for MAE, RMSE and $R^2$, respectively. A comparative chart of the reported results is graphed using the Fig. 3(b). In general, it can be observed that RFR, XGB and

LGB display similar performances in relation to SVM. The reported MAE scores, as outputted by the other models (i.e. ensemble tree algorithms), are estimated at 0.067 eV/atom RFR, 0.058 eV/atom XGB, and 0.044 eV/atom LGB. Moreover, Fig. 3(a) visualizes the regression fittings as determined by their respective $R^2$ values on the test set. Overall, SVM is once again confirmed as the best fitting model followed closely by LGB, XGB, and RFR, in that order. Moreover, it has been established in previous studies that the inclusion of the stability energy ($Es$) among the set of generalized features can further enhance the predictive efficiency of the formation energy ($Ef$) target [8]. This in essence is in agreement with the current results, as the impressive modeling capability is suggested to be due to the strong effect of $Es$, which can otherwise be referred to as the *energy above convex hull* parameter. It should be noted that although $Ef$ is not a direct measure of perovskite's stability or instability, a strong relationship is suggested to exist between $Ef$ and $Es$, which is due to the method used in thermodynamically estimating both properties, as applicable to first-principles simulations and laboratory experiments [33]. Hence, the present study points to this advantage and confirms the contributing effect of $Es$ in predicting the formation energy, if considered among the set of input features. On comparing the results obtained in the current study to previous benchmark investigations (i.e. Table 1) further authenticates the validity of the prepared perovskite dataset in the present study. For instance, the predictive performance in the present evaluation on $Ef$ is demonstrated to improve the modeling accuracy by 76% when compared to a previous study on tabular-dataset modeling using SVM and a similar set of input features [8]. The better modeling performance can attributed to the dataset robustness in size and variability, as all ML algorithms are provided with more information and complexities for effective training.

Table 3. Test set error evaluation on $Ef$ and $Eg$ prediction, as outputted by all considered ML models. SVM and LGB are seen as the best predictive models for simulating $Ef$ and $Eg$, respectively.

| Model | Formation Energy, *Ef* | | | Energy Band gap, *Eg* | | |
|---|---|---|---|---|---|---|
| | MAE (eV/atom) | RMSE (eV/atom) | R2 (%) | MAE (eV) | RMSE (eV) | R2 (%) |
| **SVM** | 0.013 | 0.070 | 99.45 | 0.297 | 0.569 | 79.79 |
| **RFR** | 0.067 | 0.131 | 98.06 | 0.220 | 0.457 | 86.93 |
| **XGB** | 0.058 | 0.111 | 98.61 | 0.234 | 0.464 | 86.52 |
| **LGB** | 0.044 | 0.088 | 99.12 | 0.216 | 0.440 | 87.90 |

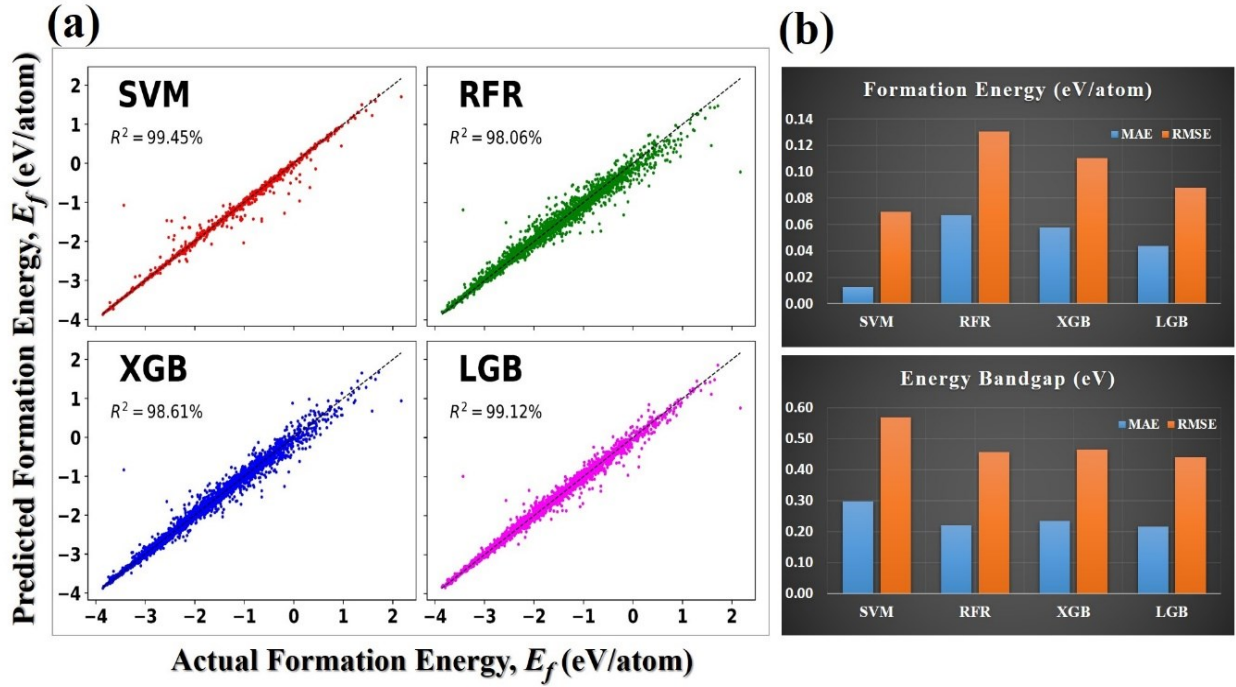

Figure 3. (a) Regression fitting performances for all models on test set; (b) A comparative chart of MAE and RMSE scores for *Ef* and *Eg* on test set.

## 5.2. Energy band gap (*Eg*) prediction

For predicting the energy band gap, the dataset is reorganized into a $M \times N = 16{,}323$ (perovskites) $\times$ 59 (features) matrix, and is split into 70% and 30% for training and testing, respectively. The input features used for simulation comprise of all features as applicable to the prediction of the formation energy (*Ef*) while also including *Ef* as an additional input variable. As demonstrated in past studies, including *Ef* among the set of input features is known to be effective in improving the modeling accuracy of band gap prediction [3, 8]. As reported using the Table 3, the LGB model is demonstrated to outperform its peers in the prediction of *Eg* based on all evaluation metrics. The reported results are comparatively graphed in the Fig. 3(b) and the evaluated MAE scores for all models are outputted at 0.297 eV, 0.220 eV, 0.234 eV and 0.216 eV for SVM, RFR, XGB, and LGB, respectively. As a result, the current evaluation underlines ensemble tree algorithms (i.e. LGB, XGB and RFR) as superior models for band gap prediction. Besides, on comparing the current band gap predictive accuracy to benchmark previous studies by Chenebuah et al. [8] and Li et al. [3] translates into about 53% and 44% in MAE improvement accuracy levels, respectively. In general however, higher marginal errors can be considerably observed with band gap evaluation unlike their other target properties counterparts. This is suggested to be due to the problem of band gap underestimation using DFT-simulation [34]. Moreover, the current dataset is highly generalized by considering both metallic/infinite band gaps (i.e. $Eg = 0$) and non-metallic/finite band gaps (i.e. $Eg > 0$) among the set of data examples. The lower precision in outputted band gap scores therefore suggests the need for further investigation using improvised features that holistically capture bandgap quantum mechanical behavior. This is an area of future study.

### 5.3. Crystal system (*Cs*) multi-classification

On final investigation, multi-classification experiments are implemented on the current dataset for the assignments of crystal systems to their respective classes. As illustrated in Fig. 2(b), the dataset is seen to be highly imbalanced with cubic, trigonal, orthorhombic and tetragonal systems, providing over 94% of all data samples. On the other hand, monoclinic, hexagonal and triclinic structures are negligibly represented due to their much lower data sample instances. In order to avoid the skewing of the classification exercise to the majority classes, these three classes with negligible sample sizes are excluded from the analysis. As such, excluding the dataset from these negligible crystal system entries results into 15,413 structures. In addition, the multi-classification experiment is investigated based on downsampling and oversampling techniques. For the case of downsampling, the dataset is randomly downsized to produce equal examples for each class that are representative of the lowest under-represented crystal system, which in this case is the tetragonal crystal class with 2089 samples. Therefore, cubic, trigonal and orthorhombic systems are all downsampled to yield 2089 $ABX_3$ compounds each. As a result, the dataset is reorganized into a $M \times N = 8{,}356$ (perovskites) $\times$ 58 (features) matrix, and is split into 70% and 30% for training and testing, respectively. Moreover, the input features used in the multi-classification experiment are selected as the same set of features that were previously used in the prediction of the formation energy. For machine-readable purposes, each crystal system target is uniquely encoded using discrete/categorical representative forms. As reported using the Table 4, the multi-classification experiment is demonstrated to output average F1 accuracy at 0.85 each for SVC, XGB and LGB models, implying that about 85% of all considered samples were correctly assigned to their respective crystal classes. For the RFR model however, a slightly lower accuracy at 82% is achieved. Besides, on taking a closer look into the modeling behavior indicates that trigonal and orthorhombic systems are reported with good Precision, Recall, and F1-scores, which are all above 90% for all considered models. The main challenge however, is in distinguishing between cubic and tetragonal structures, as highlighted using the Table 4 and further illustrated using confusion matrices in Fig. 4. According to solid-state crystallography, cubic, tetragonal and orthorhombic systems can be regarded as structurally orthogonal, which means that all three inter-axial angles in $\alpha$, $\beta$, and $\gamma$ are 90º in their conventional unit cell but with different lattice edge-vector features [35]. As such, the inherent geometrical or volumetric overlaps between cubic and tetragonal is suggested to be the primary reason for the relatively lower classification results, as the present descriptor set used in modeling the *Cs* target show significant deficiency in distinguishing between both systems.

Table 4. Multi-classification report from the random downsampling experiment, as evaluated on test set data entries for all considered ML models. To avoid the biased adjudication of the ML model to a majority class, note that each crystal system yields 30% of their respective sample sizes to their corresponding test sets.

| Crystal System | SVC | | | RFR | | | XGB | | | LGB | | | Support |
|---|---|---|---|---|---|---|---|---|---|---|---|---|---|
| | Pre | Rec | F1 | Pre | Rec | F1 | Pre | Rec | F1 | Pre | Rec | F1 | |
| **Cubic** | 0.77 | 0.72 | 0.74 | 0.70 | 0.70 | 0.70 | 0.76 | 0.75 | 0.75 | 0.76 | 0.75 | 0.75 | 627 |
| **Trigonal** | 0.95 | 0.95 | 0.95 | 0.94 | 0.94 | 0.94 | 0.95 | 0.96 | 0.96 | 0.96 | 0.96 | 0.96 | 627 |
| **Orthorhombic** | 0.95 | 0.94 | 0.94 | 0.92 | 0.95 | 0.93 | 0.95 | 0.95 | 0.95 | 0.95 | 0.96 | 0.95 | 627 |
| **Tetragonal** | 0.73 | 0.78 | 0.75 | 0.70 | 0.69 | 0.69 | 0.75 | 0.75 | 0.75 | 0.74 | 0.74 | 0.74 | 627 |
| | | | | | | | | | | | | | |
| **Accuracy** | 0.85 | | | 0.82 | | | 0.85 | | | 0.85 | | | 2508 |

\* Pre and Rec are Precision and Recall metrics, respectively.

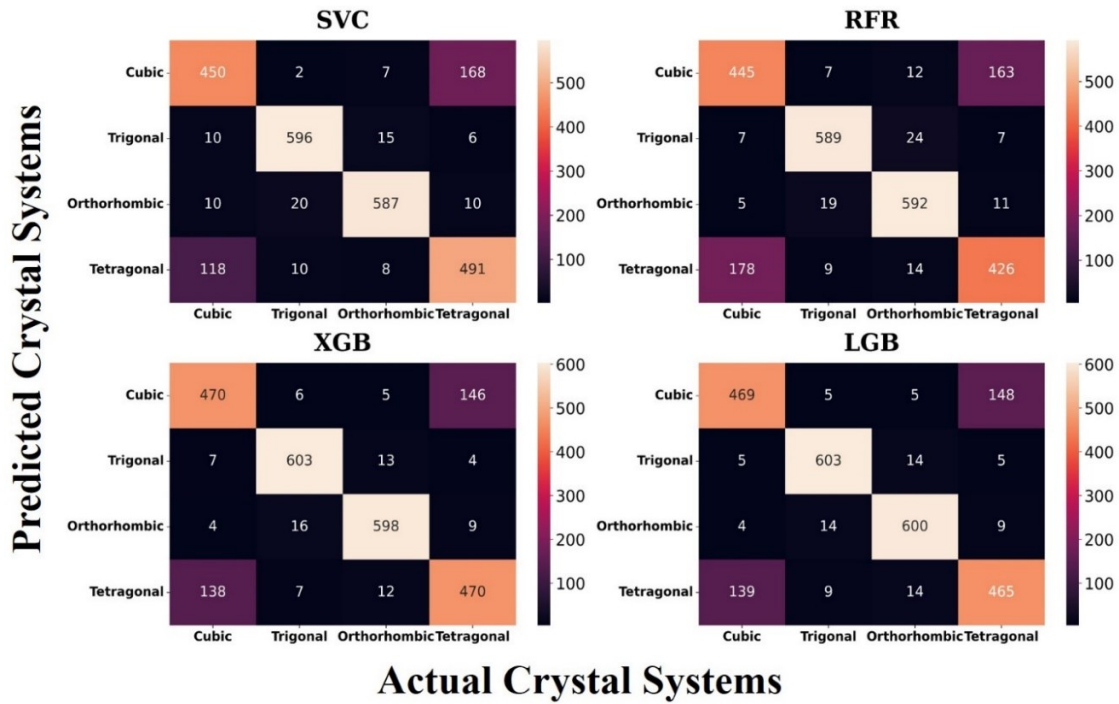

Figure 4. Heat map of confusion matrices for all considered models, as outputted based on the test set report.

To further highlight on the modeling intricacies of the multi-classification experiment with respect to data augmentation, oversampling technique is investigated on all considered under-represented crystal systems using the Synthetic Minority Over-Sampling Technique (SMOTE) [36]. In general, SMOTE creates synthetic data points by interpolating between existing minority data points along the lines of *k*-nearest neighbors in hyperdimensional feature space. As a result, minority classes in trigonal, orthorhombic and tetragonal crystal systems are each oversampled using SMOTE to reproduce new data examples that are of equal size to their cubic training data counterpart (i.e. until a maximum of 3,881 samples are reached). Moreover, the task with oversampling is conducted only on the training data, ensuring that the original samples subjected to testing are uncompromised (i.e. original and not synthesized). Emerging from the SMOTE investigation is the Fig. 5, which graphically compares the F1 scores between downsampling and oversampling experiments with respect to each crystal system and ML models. In general, it can thus be observed that the effect of oversampling is revealed to have little to no significant impact on the *overall Accuracy* F1 scores for all models. In specific to a crystal system however, the average F1-score realized with the cubic crystal system is shown to improve across all models, but at the expense of the tetragonal system.

This suggests that the SMOTE process detrimentally interprets tetragonal systems to be cubic during the synthetic generation process. For further highlighting on the difficulty in distinguishing between cubic and tetragonal, Fig. 6 illustrates Principal Components (PC) using the top-three factorial axes that best translates the largest variance. For both downsampling and oversampling cases, the clustering in data distribution is seemingly interwoven for cubic and tetragonal. On comparing cases with and without oversampling experimentation, the marginal increase in F1-score for cubic is estimated at 7%, 10%, 9%, and 9%, for SVM, RFR, LGB, and XGB, respectively. Likewise, the marginal decrease in F1-scores for tetragonal is estimated at 28%, 39%, 24%, and 24%. However, orthorhombic and trigonal systems maintain their scores and are considerably unaffected by the oversampling process. This in agreement with the PC cluster plots as orthorhombic and trigonal systems are seen to clearly distinguishable in hyperdimension. Generally, the average accuracy scores obtained from the multi-classification experiment affirms the resourcefulness of the present dataset for benchmark ML analysis. For instance, comparing the overall F1 result obtained in the present study to the multi-classification experiment by Behara et al. on $ABO_3$ compounds [9] demonstrates about 6% in improvement accuracy, based on the considered crystal classes.

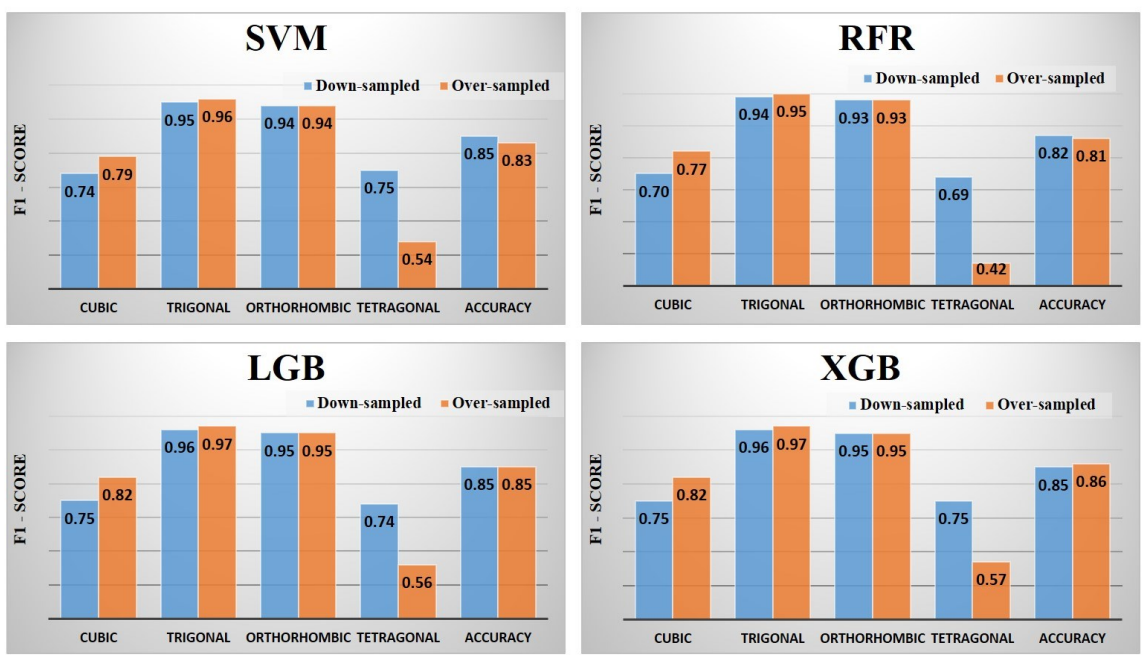

Figure 5. Comparative graph of F1-scores in the multi-classification of crystal systems based on data augmentation (oversampling) for all considered ML models.

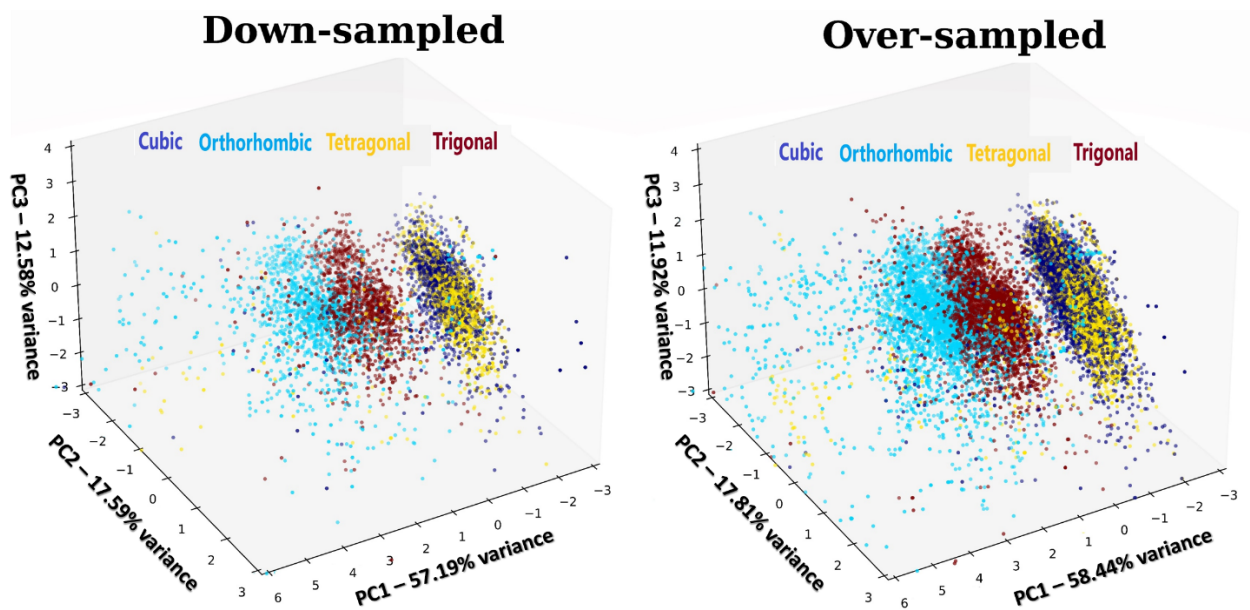

Figure 6. Principal Component (PC) cluster plots revealing the top-three axes that best captures variance in targeted data distribution with respect to downsampling and oversampling techniques, as it relates to the multi-classification of crystal systems.

## 6. Conclusion

In this study, a robust tabular-dataset of 16,323 $ABX_3$ perovskite structures is extracted and preprocessed for benchmark machine learning investigation. The dataset is developed using sixty-one generalized input features, which are broadly divided into three categories, namely: physicochemical, stability/geometrical, target electronic properties. The target properties subjected to benchmark evaluation includes formation energy prediction, energy band gap prediction and crystal system multi-classification. Four ML models are used in the experiment and were chosen based on their previously reported accuracy on tabular datasets for general inorganic solid-state ML analysis. They include Support Vector Machine (SVM), Random Forest (RFR), eXtreme Gradient Boosting (XGB), and Light Gradient Boosting (LGB) algorithms. The best accuracy realized in study are reported at 0.013 eV/atom MAE, 0.216 eV MAE, and 85% F1, corresponding to formation energy prediction, band gap prediction and crystal system multi-classification, respectively, which represents considerable improvement when compared to previous benchmark evaluation. The good accuracy in modeling behavior thereby authenticates and validates the current dataset for benchmark ML usage.

The preprocessed dataset is made openly available as resource for enhancing target predictive capability via ML in the field of materials informatics. Future studies will focus on further advancing the modeling accuracy of target electronic properties by using deep generative ML models.

**Data availability**

The raw data required to reproduce these findings are directly available to download from www.oqmd.org. The OQMD RESTful API was used to generically search for exclusive $ABX_3$ structures. The processed dataset and Python source codes required to reproduce these findings are available to download from github.com/chenebuah/ML_abx3_dataset